\documentclass[
 amsmath,amssymb,
 aps,
]{revtex4-2}

\usepackage{graphicx}
\usepackage{dcolumn}
\usepackage{bm}
\usepackage{amsmath}
\usepackage{slashed}
\usepackage{revsymb}

\makeatletter
\newcommand\makebig[2]{%
  \@xp\newcommand\@xp*\csname#1\endcsname{\bBigg@{#2}}%
  \@xp\newcommand\@xp*\csname#1l\endcsname{\@xp\mathopen\csname#1\endcsname}%
  \@xp\newcommand\@xp*\csname#1r\endcsname{\@xp\mathclose\csname#1\endcsname}%
}
\makeatother

\makebig{biggg} {4.0}
\makebig{Biggg} {4.5}
\makebig{bigggg}{5.0}
\makebig{Bigggg}{5.5}

\begin{document}

\preprint{APS/123-QED}

\title{Electron wave spin in a cavity}

\author{Ju Gao}
\email{jugao@illinois.edu}
\author{Fang Shen}%
\affiliation{University of Illinois, Department of Electrical and Computer Engineering, Urbana, 61801, USA}




\date{\today}

\begin{abstract}
Our study reveals electron spin in a cavity as a stable circulating current density, characterized by a torus topology. This current density circulates concentrically beyond the cavity boundary, illustrating the concept of evanescent wave spin. While the interaction with a uniform magnetic field aligns with established spin-field observations, our analysis of regional contributions deviates from particle-based spin predictions. The integration of charge and spin properties into a single Lorentz covariant entity suggests that the electron wave constitutes the fundamental and deterministic reality of the electron.

\end{abstract}

\pacs{3.50, 32.80, 42.50}

\maketitle


\section{\label{sec:Intro}Electron Wave Spin}
This paper constitutes a continuous investigation of the electron wave spin picture~\citep{Belinfante39, Ohanian86, Sebens2019, GaoJOPCO22}, which interprets electron spin as circulating current or momentum densities derived as observable quantities from the Dirac theorem. The combination of current density with charge density yields a Lorentz covariant and charge-conserved electron wave, inherently aligning with special relativity. Notably, this view mirrors the four-current description of electricity in classical electromagnetism~\citep{JDJackson1999}, presenting a Lorentz covariant and charge-conserved electromagnetic entity.

Recent investigations~\citep{GaoQeiosFrac2023} of electron spin in quantum wells have revealed wealth of topological information encoded within the electron wave spin that are absent in the conventional particle spin picture. The interaction of this current density with a magnetic field unveils an intrinsic electron spin value of $\hbar /2$, influenced by boundary geometry and quantum numbers. Furthermore, our studies~\citep{GaoQeiosEva2023} show that tunnelled electron waves outside quantum wells exhibit spin behavior alongside the wave spin inside the well. These findings underscore the non-local and global nature inherent to the electron wave, extending across its entirety.

Electron spin, a fundamental property of electrons,plays a pivotal role in the advancement of quantum physics~\citep{Stern22,Dirac28,Pauli26,Commins12, Jammer66} and related technologies~\citep{HIROHATA2020166711, Ladd10}. Initially conceptualized as the rotation of an electron particle around its axis~\citep{Uhlenbeck26}, the conventional understanding of electron spin faced challenges in reconciling with the principles of special relativity. This discrepancy led to the consideration of spin as an abstract two-valued property devoid of tangible physical significance. Nonetheless, electron spin has traditionally been perceived as a local property inherent to the electron particle. This particle spin view is fundamentally tied to the conventional wave-particle interpretation of quantum mechanics, where the electron particle is considered as the fundamental entity, with its presence described by the square of the wavefunction as a probability map or "electron cloud". Here, the electron wave, or "electron cloud", represents a statistical abstraction, not a real entity. However, this duality interpretation poses challenges in reconciling with special relativity and causality, particularly in the single electron presence of distant regions of an arbitrary wave and in entangled electrons exchanging spin information over distance.

Conversely, the wave spin picture illustrates that the electron wave, or the "electron cloud" itself, spins. The integration of the charge and spin properties into a single Lorentz covariant entity compels us to propose that the electron wave constitutes the fundamental reality of the electron, rather than a mere statistical abstraction. In this wave entity view, an electron always exists in alignment with special relativity, and entangled electrons require no "spooky" action to exchange spin information between the globally existent and overlapping waves.

This shift in viewpoint holds implications for understanding quantum mechanics and exploring novel spin effects. However, our previous studies have predominantly focused on the two-dimensional confinement of Dirac electrons, assuming weak or negligible third-dimensional confinement. To more accurately simulate scenarios akin to molecules~\cite{doi:10.1126/science.adj5328} or quantum dots~\cite{Efros2021}, our forthcoming research will delve into exploring wave spin in a three-dimensional cavity. This approach will enable a more accurate representation of real-world conditions and facilitate experimental validation of the distinctions between competing views on electron spin. 

\section{\label{sec:CavitySpinCylinder}Electron Wave spin in a cavity}
To investigate the electron wave spin within a cavity, we begin with solving the Dirac equation~\citep{Dirac28}:
\begin{equation} \label{Dirac}
i\hbar (\partial /\partial t)\Psi (\pmb{r},t)=\left[ -i\hbar c \pmb{\alpha} \cdot \pmb{\nabla }+\gamma ^0 \text{mc}^2+U(\pmb{r})\right]\Psi(\pmb{r},t)
\end{equation}
where $m$ and $c$ are the electron mass and the speed of light, respectively. $ \hbar $ denotes the reduced Planck constant. 

The operator in the cylindrical coordinate is expressed as:
\begin{equation}
\pmb{\alpha} \cdot \pmb{\nabla}
=\alpha _{\rho }\frac{\partial }{\partial \rho }+\alpha _{\phi }\frac{1}{\rho }
\frac{\partial }{\partial \phi }+\alpha _z
\frac{\partial }{\partial z},
\end{equation}
where $(\rho,\phi,z)$ represent polar, azimuthal angle and z coordinate, respectively. The $\alpha-$matrix in the cylindrical coordinate follows the normalization and commuting properties:
\begin{eqnarray}
&&\sigma _{\rho }^2=\sigma _{\phi }^2=\sigma _{z}^2=1, \nonumber \\
&&[\sigma _{\rho },\sigma _{\phi }]=2i \alpha _z, \nonumber \\
&&\{ \sigma _{\rho },\sigma _{\phi }\}=0. \nonumber \\
\end{eqnarray}

The potential $U(\pmb{r})$ in the Dirac equation represents a cylindrical cavity with radius $R$ and height $2d$, 
\begin{equation}\label{potential}
U(\pmb{r})=\Biggl\{
\begin{array}{ccc}
0, & 0<\rho<R;-d<z<d & \text{Region I}\\
U, & \rho>R;-d<z<d & \text{Region II}\\
\infty,& z<-d; z>d & \text{Region III} \\
\end{array}
\end{equation}
where $U$ corresponds to a finite potential, typically on the order of a few or a fraction of an electronvolt (eV).

Within this system, the electron is completely confined within the Regions I and II due to an infinite potential barrier along $z$-axis. However, in the radial direction within the planar structure, the electron wave is not fully isolated. Consequently, the electron wave can tunnel beyond the finite radial barrier, making it a subject of investigations. Although this potential model effectively describes planar quantum devices, it has been employed only for solving the Schr$\ddot{o}$dinger equation~\citep{lobanova2004cylindrical}. Rigorous solution of the Dirac equation in this potential becomes our central objective in order to investigate the wave spin. 

In this section, we present only the main results of our solution, reserving the detailed derivation procedures for the Appendix. This arrangement allows us to focus the discussion on the wave spin topology and interaction behaviour in the main text. 

Before obtaining wavefunction solutions in the cavity, we assume the wavefunction in Region III to be zero. However, in reality, the potential cannot be infinite, resulting in the existence of an evanescent electron wave in Region III, as previously discussed~\cite{GaoQeiosEva2023}. Remarkably, even for an infinitely large potential, we have demonstrated that the evanescent wave persists within the skin-depth range at the boundary. Meanwhile, the wavefunction inside Regions I and II can be accurately solved by imposing a zero wavefunction boundary condition for the large component wavefunction.

The wavefunction within Regions I and II can be expressed as follows:
\begin{equation}
\Psi (\pmb{r},t)=e^{-i\mathcal{E}t/\hbar}
\psi (\rho,\phi,z),
\end{equation}
where $\mathcal{E}$ represents the eigenenergy, given that the potential within the cavity is time-independent.

After following the detailed procedures outlined in the Appendix, we present the wavefunction expression for the spin-up Dirac electron within the cavity:
\begin{widetext}
\begin{equation}\label{psinlm}
\psi_{nlm\uparrow}(\rho,\phi,z)=\Biggg\{
\begin{array}{cc}e^{il\phi} N \left(
\begin{array}{c}
J_l(\zeta_{nlm}  \rho )\cos(k_m z) \\
 0 \\
-i \eta_{I} k_m J_l(\zeta_{nlm}  \rho )\sin(k_m z) \\
i e^{i \phi}\eta_{I} \{\frac{\zeta_{nlm}}{2}\left[ J_{l-1}(\zeta_{nlm}  \rho )-J_{l+1}(\zeta_{nlm}  \rho )\right]-\frac{l}{\rho } J_l(\zeta_{nlm}  \rho )\}\cos(k_m z) \\
\end{array}
\right)  & \text{Region I}\\\\
\kappa e^{il\phi} N \left(
\begin{array}{c}
K_l(\xi_{nlm} \rho)\cos(k_m z) \\
 0 \\
-i \eta_{II} k_m K_l(\xi_{nlm}  \rho )\sin(k_m z) \\
i e^{i \phi}\eta_{II} \{\frac{\xi_{nlm}}{2}\left[-K_{l-1}(\xi_{nlm} \rho)-K_{l+1}(\xi_{nlm} \rho)\right]-\frac{l}{\rho } K_l(\xi_{nlm} \rho)\}\cos(k_m z)  \\
\end{array}
\right)  & \text{Region 1I}
\end{array}
\end{equation}
\end{widetext}
where $N$ is the normalization constant. $ \kappa$ is a constant associated with the wavefunction in Region II, which is state-specific to ensure satisfaction of the boundary conditions.

In Eq.~\ref{psinlm}, the term $e^{il\phi}$ represents the azimuthal wavefunction with azimuthal quantum number $l=0,1,2...$. The $z$-coordinate wavefunction corresponds to a standing wave with the wave vector $ k_m=\frac{m \pi}{2d} $, where $ m=1,3,5... $ denotes the $ z $-axis quantum number. The radial wavefunctions are described by the Bessel functions of the first and second kinds, denoted as $J_{l}(\zeta_{nlm}  \rho )$ and $ K_l(\xi_{nlm} \rho) $, respectively. Here, $n=1,2,3...$ represents the radial quantum number. The corresponding wave vectors $\zeta_{nlm}$ and $\xi_{nlm}$ depend on the azimuthal and $z$-axis quantum numbers $l$ and $m$, while their implicit  dependence on the radial quantum number $n$ arises from the boundary condition. Notably, $\zeta_{nlm}$ and $\xi_{nlm}$ are related to the eigenenergy $\mathcal{E}_{nlm}$ by:
\begin{eqnarray}\label{zetaxi}
\zeta_{nlm}^2&=&\frac{\mathcal{E}_{nlm}^2-m^2 c^4-\hbar ^2 c^2 k_m^2}{\hbar ^2 c^2}; \nonumber \\
\xi_{nlm}^2&=&\frac{(\mathcal{E}_{nlm}-U)^2-m^2 c^4-\hbar ^2 c^2 k_m^2}{\hbar ^2 c^2 }. \nonumber \\
\end{eqnarray}

Other factors $ \eta _I $ and $ \eta _{II} $ in Eq.~\ref{psinlm} are defined by: 
\begin{eqnarray}\label{eta}
\eta _I&=&\frac{\hbar c}{\mathcal{E}+m c^2};\nonumber\\
\eta _{II}&=&\frac{\hbar c}{\mathcal{E}-U+m c^2}, 
\end{eqnarray}
which converge to a common value:
\begin{equation}\label{eta2}
\eta _{I} \approx \eta _{II} \approx \eta=\frac{\hbar c}{2m c^2}=\frac{\hbar}{2mc},
\end{equation}
due to the fact that the energy $ \mathcal{E}$ closely approximates the electron rest energy: $ \mathcal{E}\approx m c^2$, for electrons in the cavity of our interest. 

As an example, we focus our study on the lowest quantum state of $l=0$ and $m=1$. The eigenwavefunction for the spin-up electron is now expressed as:
\begin{widetext}
\begin{equation}\label{psin01}
\psi_{n01\uparrow}(\rho,\phi,z)=\Biggg\{
\begin{array}{cc}N\left(
\begin{array}{c}
J_0(\zeta_{n01}  \rho )\cos(k z) \\
 0 \\
-i \eta k J_0(\zeta_{n01}  \rho )\sin(k z) \\
-i e^{i \phi}\eta \zeta_{n01} J_1(\zeta_{n01}  \rho )J_1(\zeta_{n01}  \rho )\cos(k z) \\
\end{array}
\right)  & \text{Region I}\\\\
\kappa N\left(
\begin{array}{c}
K_0(\xi_{n01} \rho)\cos(k z) \\
 0 \\
-i \eta k K_0(\xi_{n01}  \rho )\sin(k z) \\
-i e^{i \phi}\eta \xi_{n01} K_1(\xi_{n01} \rho)K_l(\xi_{n01} \rho)\cos(kz)  \\
\end{array}
\right)  & \text{Region 1I}
\end{array}
\end{equation}
\end{widetext}
where we set $ k_1=k=\frac{\pi}{2d} $. 

We now apply boundary conditions at $ \rho=R $ for all components of the bispinor wavefunction in Eq.~\ref{psin01}:
\begin{eqnarray}\label{boundary}
J_0(\zeta_{n01} R)&=&\kappa K_0(\xi_{n01}  R);\nonumber \\
\zeta_{n01} J_1(\zeta_{n01} R)&=& \kappa \xi_{n01} K_1(\xi_{n01} R).
\end{eqnarray}
which determine the eigenenergy $\mathcal{E}_{n01}$ and $\kappa  $ by using the expressions from Eqs.~\ref{zetaxi}. 

Up to this point, we have derived the full eigen solution for the Dirac electron in the cavity. This result will facilitate scientific investigations and support engineering exploration of wave spin behavior in confined environments.

\section{\label{sec:Topology}Torus topology of Electron Wave spin in a cavity}
In classic electromagnetism, the electric four-current density-comprising both charge density $q(\pmb r)$ and current density $\pmb j(\pmb r)$-plays a fundamental role in determining nearly all electromagnetic phenomena~\cite{JDJackson1999}. This four-current is Lorentz covariant, represents a unified electromagnetic entity. Investigating the four-current for quantum objects is both natural and necessary, as none of the principles of quantum mechanics contradict the basic laws of electromagnetism and relativity.

In quantum mechanics, the Lorentz covariant four current of charge and current density are observables that can be calculated from the wavefunctions derived in Sec.~\ref{sec:CavitySpinCylinder}:
\begin{eqnarray}\label{fourcurrent}
q(\rho ,\phi ,z)&=&e\psi_{n01\uparrow}^{\dagger}(\rho,\phi,z)\psi_{n01\uparrow}(\rho,\phi,z); \nonumber \\
\pmb{j}(\rho ,\phi ,z)&=&ec\psi_{n01\uparrow}^{\dagger}(\rho,\phi,z)\pmb{\alpha}\psi_{n01\uparrow}(\rho,\phi,z),
\end{eqnarray}
where $e=1.602\times 10^{-19}\text{C}$ represents the elementary charge. The charge and current densities adhere to charge conservation principles throughout. 

Utilizing the wavefunction expression in Eq.~\ref{psin01}, we arrive at the following explicit expressions:
\begin{eqnarray}\label{fourcurrent2}
q(\rho,z)&=&\Biggl\{
\begin{array}{cc}
eN^2 J_0(\zeta_{n01} \rho)^2\cos^2 (k z), \text{Region I}\\\\
eN^2 \kappa^2 K_0(\xi_{n01} \rho)^2 \cos^2 (k z), \text{Region II}\\
\end{array};\nonumber \\
j_\phi(\rho,z)&=&\Biggl\{
\begin{array}{cc}
-2 N^2 ec \eta \zeta_{n01} J_0(\zeta_{n01} \rho) J_1(\zeta_{n01} \rho)\cos^2 (k z), \text{Region I}\\\\
-2 N^2 \kappa^2 ec \eta \xi_{n01} K_0(\xi_{n01} \rho) K_1(\xi_{n01} \rho)\cos^2 (k z), \text{Region II}\\ 
\end{array}. \nonumber \\
j_\rho(\rho,z)&=&0, \, \text{everywhere} \nonumber \\
j_z(\rho,z)&=&0, \, \text{everywhere} \nonumber \\
\end{eqnarray}

Here, we drop the $\eta^2$ term in the charge expression. This is justified by the fact that $\eta$ is on the order of the Compton wavelength, which is significantly smaller than the cavity size of interest.

The presence of a solitary azimuthal current density underscores the wave-like nature of the electron spin. This current exists globally and circulates concentrically both within and beyond the cavity, confirming the evanescent wave spin previously disclosed in the two-dimensional confinement. 

We now calculate the normalization factor by integrating the charge density across all regions to yield unit electron charge: 
\begin{eqnarray}\label{norm}
e&=&\int _0^{\infty }\int _0^{2 \pi }\int _{-d}^d q(\rho,z)\rho d\rho d\phi dz \nonumber \\
&=&e N^2 2 \pi d \left[\int_0^R J_0 (\zeta_{n01}  \rho )^2 \rho d\rho +\kappa ^2 \int_R^{\infty }K_0 (\xi_{n01}  \rho )^2 \rho d\rho \right] \nonumber \\
\end{eqnarray}
which gives
\begin{equation}\label{N2}
N^2=\frac{1}{2\pi d \left[\int_0^R J_0 (\zeta_{n01}  \rho )^2 \rho d\rho +\kappa ^2 \int_R^{\infty }K_0 (\xi_{n01}  \rho )^2 \rho d\rho \right]}. 
\end{equation}

While an analytical expression for the integral in Eq.~\ref{N2} exists, we intentionally retain the integral form to elaborate on the contributions from both the confined and evanescent regions.

To gain insight into the electron wave spin behavior, let us consider a realistic quantum dot cavity with the following dimensions: $R=8 \text{nm}$ (radius) and $d=4\text{nm}$ (height). The potential is set to $U=10 \text{meV}$.

Applying the boundary condition from Eq.~\ref{boundary}, we determine the ground state eigenenergy as $\mathcal{E}_{101}-m c^2=8.06 \text{meV}$ and the wavefunction coupling constant $\kappa=15.9$. Subsequently, we calculate the wave vectors $\zeta_{101}=2.40 \times 10^8\text m^{-1}$ and $\xi_{101}=4.53 \times 10^8\text m^{-1}$ using Eqs.~\ref{zetaxi}. These parameters allow us to fully describe the charge and current densities numerically and graphically.

\begin{figure}
\includegraphics[width=0.4\textwidth]{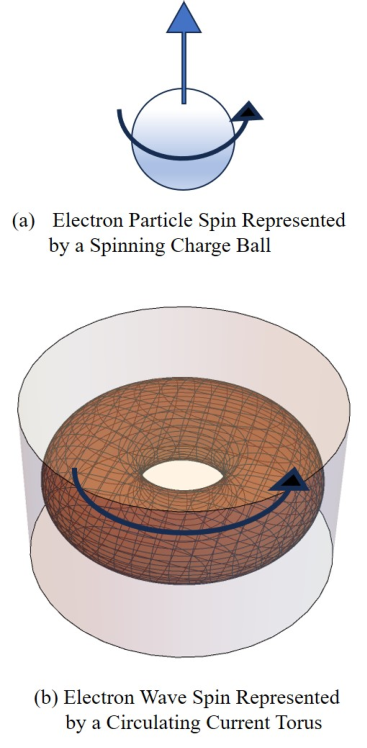}%
\caption{\label{fig:Fig1} Figure (a) illustrates the particle spin representation of a rotating corpuscular electron. In contrast, Figure (b) depicts the wave spin view of an electron in the ground state $(101)$, with an eigenenergy of $\mathcal{E}_{101}-m c^2=8.06 \text meV$. This electron is confined within a cavity characterized by a radius  $R=8 \text{nm}$, height $2d=8\text{nm}$, and potential energy $U=10 \text{meV}$. The ring torus topology emerges from the three-dimensional contour plot of the current density at the $2/3$ contour peak level.}
\end{figure}

We now choose to employ a three-dimensional contour plot to visualize the current density in Fig.~\ref{fig:Fig1}. 
The electron wave spin exhibits a ring torus topology, which remains entirely confined within the cavity at $2/3$ contour peak level. The torus topology complements the donut topology in the two-dimensional quantum well, with the additional confinement along the $z$-axis. The torus topology of the electron wave spin fundamentally differs from that of particle spin. We thus anticipate that this alternative perspective on electron spin will stimulate discussions across various scientific domains, considering that topological features are often well-represented and preserved.    

\section{\label{sec:Detection}Electron wave spin interaction with a field}
The wave spin picture, characterized by the current density, offers a conceptually and topologically distinct perspective on electron spin. Explicitly derived expressions for current density serve as a testing ground for experimental verification and exploration of new spin effects. In this section, we investigate the interaction of wave spin with an external field, since the spin is fundamentally observed and measured through the interaction $\pmb{j}(\pmb r)\cdot \pmb{A}(\pmb r)$, where $\pmb{A}(\pmb r)$ represents the vector potential associated with the external field.

To compare with existing experimental results, we choose a magnetic vector potential $\pmb{A}(\pmb r)=(0,A_{\phi},0)$, where
\begin{equation}\label{A}
A_{\phi}(\rho) = \frac{B}{2} \rho,
\end{equation}
to produce a uniform magnetic field $B$ along the $z$-axis, as governed by the relation $ \pmb{B}= \nabla \times \pmb{A}$. 

We proceed to calculate the interaction energy for the ground state electron $(101)$ illustrated in Fig.~\ref{fig:Fig1} by integrating 
$ \pmb{j}\cdot \pmb{A} $ across all relevant regions:
\begin{widetext}
\begin{eqnarray}\label{intwave}
\mathcal{E}_\text{int}^{wav}&=&\int _{0}^{\infty}\int _{0}^{2\pi}\int _{-d}^{d} j_\phi(\rho,z) A_\phi(\rho) \rho d\rho d\phi dz \nonumber \\
&=& N^2 e c \eta B 2 \pi d \left[\int_0^R  \zeta_{101}  J_0(\zeta_{101} \rho ) J_1(\zeta_{101} \rho )\rho ^2\, d\rho +\kappa ^2 \int_R^{\infty }  \xi_{101} K_0(\xi_{101} \rho ) K_1(\xi_{101} \rho )\rho ^2 \, d\rho \right] \nonumber \\
&=&\frac{e \hbar }{2m} B \frac{\int_0^R  \zeta_{101}  J_0(\zeta_{101} \rho ) J_1(\zeta_{101} \rho )\rho ^2\, d\rho +\kappa ^2 \int_R^{\infty }  \xi_{101} K_0(\xi_{101} \rho ) K_1(\xi_{101} \rho )\rho ^2 \, d\rho}{\int_0^R J_0 (\zeta_{101}  \rho )^2 \rho d\rho +\kappa ^2 \int_R^{\infty }K_0 (\xi_{101}  \rho )^2 \rho d\rho} \nonumber \\
&=&\mu_B B 
\end{eqnarray}
\end{widetext}
where we use the definition of the Bohr magnetic moment:
\begin{equation}
\mu_B=\frac{e \hbar }{2m}, 
\end{equation}
and the relation
\begin{equation}\label{unity}
\frac{\int_0^R  \zeta_{101}  J_0(\zeta_{101} \rho ) J_1(\zeta_{101} \rho )\rho ^2\, d\rho +\kappa ^2 \int_R^{\infty }  \xi_{101} K_0(\xi_{101} \rho ) K_1(\xi_{101} \rho )\rho ^2 \, d\rho}{\int_0^R J_0 (\zeta_{101}  \rho )^2 \rho d\rho +\kappa ^2 \int_R^{\infty }K_0 (\xi_{101}  \rho )^2 \rho d\rho} =1 
\end{equation}
under the boundary condition specified by Eqs.~\ref{boundary}. 

The interaction energy $ \mathcal{E}_\text{int}^{wav}=\mu_B B  $ aligns with experimental observations, and predictions based on the particle spin model, where a free electron possesses an intrinsic magnetic moment $ \mu_B $. While the same result is obtained from different perspectives, it is essential to recognize that the assumption of a uniform magnetic field tends to obscure the differences between these viewpoints, since the uniform magnetic field is infinitely larger than either the cavity and the point-like particle. When the field size becomes comparable to the cavity, the regional contributions to the interaction diverge. 

To highlight the differences in the regional interaction contribution, we separately calculate the interactions for Regions I and II from the wave spin perspective: 
\begin{equation}
\mathcal{E}_\text{int}^{wav}=\Biggl\{
\begin{array}{cc}
\mu_B B \frac{\int_0^R  \zeta_{101} J_0(\zeta_{101} \rho ) J_1(\zeta_{101} \rho )\rho ^2\, d\rho}{\int_0^R J_0 (\zeta_{101}  \rho )^2 \rho d\rho +\kappa ^2 \int_R^{\infty }K_0 (\xi_{101}  \rho )^2 \rho d\rho}=0.71\mu_B B, \text{Region I}\\\\
\mu_B B \frac{\kappa ^2 \int_R^{\infty }  \xi_{101} K_0(\xi_{101} \rho ) K_1(\xi_{101} \rho )\rho ^2 \, d\rho}{\int_0^R J_0 (\zeta_{101}  \rho )^2 \rho d\rho +\kappa ^2 \int_R^{\infty }K_0 (\xi_{101}  \rho )^2 \rho d\rho}=0.29\mu_B B, \text{Region II}\\
\end{array}
\end{equation}

The result indicates that if we confine the uniform magnetic field exclusively into the interior of the cavity, only a fraction of $71\%$ spin can be observed and $29\%$ spin arises from the evanescent wave that lies outside the cavity.

By contrast, we calculate the regional interactions for Regions I and II from the particle spin perspective. In the particle spin view, the presence of the electron is described by the probability distribution:   
\begin{eqnarray}\label{probability}
P(\rho ,z)&=&\psi_{n01\uparrow}^{\dagger}(\rho,\phi,z)\psi_{n01\uparrow}(\rho,\phi,z)=q(\rho ,z)/e\nonumber \\
&=&\Biggl\{
\begin{array}{cc}
N^2 J_0(\zeta_{101} \rho)^2\cos^2 (k z), \text{Region I}\\\\
N^2 \kappa^2 K_0(\xi_{101} \rho)^2 \cos^2 (k z).  \text{Region II}\\
\end{array}
\end{eqnarray}

Integrating this probability density across all relevant regions yields the total interaction energy: 
\begin{eqnarray}
\mathcal{E}_\text{int}^{par}&=&\int _0^{\infty }\int _0^{2 \pi }\int _{-d}^d\mu _{B} B P(\rho ,z)\rho d\rho d\phi dz \nonumber \\
&=&\mu _B B \frac{\int_0^R J_0 (\zeta_{101}  \rho )^2 \rho d\rho +\kappa ^2 \int_R^{\infty }K_0 (\xi_{101}  \rho )^2 \rho d\rho}{\int_0^R J_0 (\zeta_{101} \rho )^2 \rho d\rho +\kappa ^2 \int_R^{\infty }K_0 (\xi_{101}  \rho )^2 \rho d\rho}  \nonumber \\
&=&\mu _B B. 
\end{eqnarray}

Similarly, we calculate the regional contribution to the interaction from the particle spin perspective:
\begin{equation}
\mathcal{E}_\text{int}^{par}=\Biggl\{
\begin{array}{cc}
\mu_B B \frac{\int_0^R J_0 (\zeta_{101}  \rho )^2 \rho d\rho }{\int_0^R J_0 (\zeta_{101}  \rho )^2 \rho d\rho +\kappa ^2 \int_R^{\infty }K_0 (\xi_{101}  \rho )^2 \rho d\rho}=0.85\mu_B B, \text{Region I}\\\\
\mu_B B \frac{\kappa ^2 \int_R^{\infty }K_0 (\xi_{101}  \rho )^2 \rho d\rho}{\int_0^R J_0 (\zeta_{101}  \rho )^2 \rho d\rho +\kappa ^2 \int_R^{\infty }K_0 (\xi_{101}  \rho )^2 \rho d\rho}=0.15\mu_B B, \text{Region II}\\
\end{array}
\end{equation}

\begin{figure}
\includegraphics[width=0.48\textwidth]{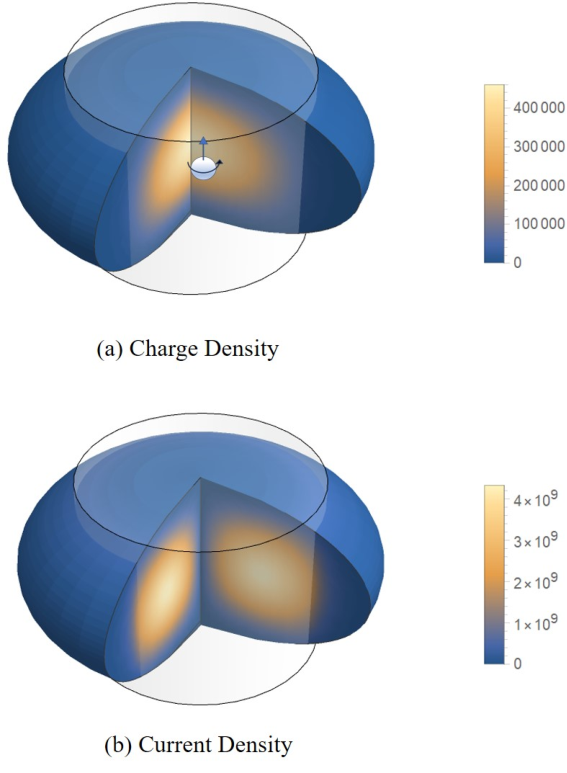}%
\caption{\label{fig:Fig2} Figure (a) shows the sliced charge density of the ground state $(101)$ electron, which is concentrated at the center of the cavity with a radius of $R=8\text{nm}$, a height of $2d=8\text{nm}$ and a potential of $U=10\text{meV}$. The particle spin model is inserted to illustrate the conventional probabilistic wave-particle view of the electron spin. In Figure (b), the sliced current density of the same ground state $(101)$ electron is displayed, aligning with the ring torus topology illustrated in Fig.~\ref{fig:Fig1}. The combined charge and current densities describe a Lorentz covariant and charge-conserved electron wave entity.}
\end{figure}

Our analysis reveals that the wave spin perspective yields almost twice the contribution to the spin-field interaction from the wave outside the cavity compared to the particle spin perspective. This phenomenon arises due to the more extended distribution of current density relative to charge density, as visually depicted in Fig.~\ref{fig:Fig2}. In the wave spin picture, spin is a global property of the electron wave, and the distribution of spin-field interaction depends on the current density. Conversely, in the particle spin picture, spin is a local property of the electron particle and the interaction distribution aligns with the probability density-an analogy to the charge density. Figure~\ref{fig:Fig2} presents a three-dimensional density plot illustrating both charge and current densities. Notably, the density plot of current density is topologically  consistent with the contour plot in Fig.~\ref{fig:Fig1}.

The analysis reveals that the two fundamentally distinct viewpoints regarding electron spin can be quantitatively distinguished. If the magnetic field were predominantly confined either inside or outside the cavity, discrepancies in interaction values would arise between the contrasting wave and particle perspectives. Leveraging advancements in microelectronic technology to engineer confined fields comparable in size to the cavity in this study allows us to experimentally validate our predictions.

\section{\label{sec:Diss}Deterministic Electron Wave Entity}
Our investigation into wave spin in a cavity reveals a ring torus topology of stable current density circulating concentrically both inside and outside the cavity. Analysing the regional spin interaction with a uniform magnetic field shows numerical discrepancies between the wave spin and particle spin predictions, providing a measurable and quantifiable means to differentiate between competing views on electron spin.

The characterization of electron spin through current density, combined with charge density, provides a unified and Lorentz covariant description of the electron wave. We propose that the electron wave, rather than the electron particle, constitutes the fundamental reality of the electron. In this perspective, the electron "particle" is understood solely as the quantized properties of the wave in charge, spin, and mass, while the corpuscular interpretation of the electron is only recognized by the wave interaction cross-section. Meanwhile, the charge and current characterizations of the wave entity are determined by wavefunctions, which are deterministic vectors in Hilbert space. Consequently, the electron wave emerges as a unified, deterministic entity for the electron, fully described by the Dirac equation and requiring no unknown parameters such as the electron radius.

Recognizing the electron wave as the true essence of the electron has significant implications for quantum science and technology. This perspective affirms that at the fundamental level, a quantum device or quantum computer operates on deterministic waves with definitive geometries and topologies, rather than on probabilistic particles with elusive parameters. 

\section{\label{sec:Acknowledgement}Acknowledgement}
The authors wish to extend their gratitude to the reviewers on Qeios for their transparent and stimulating discussions, which have greatly contributed to the development of this work.

\appendix
\section{\label{App:CavitySpinCylinder}Derivation of an electron wavefunction in a cylindrical cavity}
To investigate the electron wave spin in a cavity, we begin with solving the Dirac equation:
\begin{equation} \label{A:Dirac}
i\hbar (\partial /\partial t)\Psi (\pmb{r},t)=\left[ -i\hbar c \pmb{\alpha} \cdot \pmb{\nabla }+\gamma ^0 \text{mc}^2+U(\pmb{r})\right]\Psi(\pmb{r},t)
\end{equation}
where $m$ and $c$ are the electron mass and the speed of light, respectively. $ \hbar $ denotes the reduced Planck constant. 

The operator in the cylindrical coordinate is expressed as:
\begin{equation}\label{A:operator}
\pmb{\alpha} \cdot \pmb{\nabla}
=\alpha _{\rho }\frac{\partial }{\partial \rho }+\alpha _{\phi }\frac{1}{\rho }
\frac{\partial }{\partial \phi }+\alpha _z
\frac{\partial }{\partial z},
\end{equation}
where $(\rho,\phi,z)$ represent polar, azimuthal angle and z coordinate, respectively. The $\alpha-$matrices in the cylindrical coordinate are expressed as:
\begin{eqnarray}\label{A:alpha}
\alpha _{\rho }&=&
\left(
\begin{array}{cccc}
 0 & 0 & 0 & e^{-i \phi } \\
 0 & 0 & e^{i \phi } & 0 \\
 0 & e^{-i \phi } & 0 & 0 \\
 e^{i \phi } & 0 & 0 & 0 \\
\end{array}
\right); \nonumber \\
\alpha _{\phi }&=&
\left(
\begin{array}{cccc}
 0 & 0 & 0 & -i e^{-i \phi } \\
 0 & 0 & i e^{i \phi } & 0 \\
 0 & -i e^{-i \phi } & 0 & 0 \\
 i e^{i \phi } & 0 & 0 & 0 \\
\end{array}
\right); \nonumber \\
\alpha _z &=&
\left(
\begin{array}{cccc}
 0 & 0 & 1 & 0 \\
 0 & 0 & 0 & -1 \\
 1 & 0 & 0 & 0 \\
 0 & -1 & 0 & 0 \\
\end{array}
\right),
\end{eqnarray}
and follow the normalization and commuting properties:
\begin{eqnarray}\label{A:alpharelation}
&&\sigma _{\rho }^2=\sigma _{\phi }^2=\sigma _{z}^2=1, \nonumber \\
&&[\sigma _{\rho },\sigma _{\phi }]=2i \alpha _z, \nonumber \\
&&\{ \sigma _{\rho },\sigma _{\phi }\}=0. \nonumber \\
\end{eqnarray}

The potential $U(\pmb{r})$ in the Dirac equation represents a cylindrical cavity with radius $R$ and height $2d$, 
\begin{equation}\label{A:potential}
U(\pmb{r})=\Biggl\{
\begin{array}{ccc}
0, & 0<\rho<R;-d<z<d & \text{Region I}\\
U, & \rho>R;-d<z<d & \text{Region II}\\
\infty,& z<-d; z>d & \text{Region III} \\
\end{array}
\end{equation}
where $U$ corresponds to a finite potential, typically on the order of a few or a fraction of an electronvolt (eV).

Before obtaining the wavefunction solutions within the cavity, it is important to acknowledge that we assume the wavefunction in Region III to be zero. In reality, the potential cannot be infinite, leading to the existence of an evanescent electron wave in Region III, as previously discussed~\cite{GaoQeiosEva2023}. Remarkably, even for an infinitely large potential, we have shown that the evanescent wavefunction persists within the skin-depth range at the boundary. Meanwhile, the wavefunction inside Regions I and II can be correctly solved by imposing a zero wavefunction boundary condition for the large component wavefunction.

The wavefunction within Regions I and II can now be expressed as follows:
\begin{equation}\label{A:Psi}
\Psi (\pmb{r},t)=e^{-i\mathcal{E}t/\hbar}
\psi (\rho,\phi,z),
\end{equation}
where $\mathcal{E}$ represents the eigenenergy, given that the potential within the cavity is time-independent.

The time-independent wavefunction $ \psi (\rho,\phi,z) $ in Eq.~\ref{A:Psi} is then expressed as:
\begin{equation}
\psi (\rho,\phi,z)=\left(
\begin{array}{c}
\mu_A(\rho,\phi,z)\\
\mu_B(\rho,\phi,z)\\
\end{array}
\right),
\end{equation}
where $ \mu_A(\rho,\phi,z) $ and $ \mu_B(\rho,\phi,z) $ are spinor wavefunctions known as the large and small components of the Dirac wavefunctions. 

It is important to note that the small component wavefunction $ \mu_B(\rho,\phi,z) $ is usually neglected, leading to the reduction of the Dirac equation to the Schr$ \ddot{o} $dinger equation. In our study, we will derive analytical expressions for both $ \mu_A(\rho,\phi,z) $ and $ \mu_B(\rho,\phi,z) $.

We can derive the equations for $ \mu_A(\rho,\phi,z) $ and $ \mu_B(\rho,\phi,z) $ inside Regions I and II by substituting Eq.~\ref{A:Psi} into the Dirac equation Eq.~\ref{A:Dirac} 
\begin{widetext}
\begin{eqnarray}\label{A:muAmuB}
-i \frac{\mathcal{E}-U(\pmb{r})-m c^2}{\hbar c} \mu_A(\rho,\phi,z)&=&\left(
\begin{array}{cc}
 \frac{\partial }{\partial z } & e^{-i \phi }\frac{\partial }{\partial \rho }-i e^{-i \phi }\frac{1}{\rho }\frac{\partial }{\partial \phi } \\
 e^{i \phi }\frac{\partial }{\partial \rho }+i e^{i \phi }\frac{1}{\rho }\frac{\partial }{\partial \phi }  & -\frac{\partial }{\partial z }  \\
\end{array}
\right)\mu_B(\rho,\phi,z); \nonumber \\
-i \frac{\mathcal{E}-U(\pmb{r})+m c^2}{\hbar c} \mu_B(\rho,\phi,z)&=&\left(
\begin{array}{cc}
 \frac{\partial }{\partial z }  & e^{-i \phi }\frac{\partial }{\partial \rho }-i e^{-i \phi }\frac{1}{\rho }\frac{\partial }{\partial \phi } \\
 e^{i \phi }\frac{\partial }{\partial \rho }+i e^{i \phi }\frac{1}{\rho }\frac{\partial }{\partial \phi }  & -\frac{\partial }{\partial z }  \\
\end{array}
\right)u_A(\rho,\phi,z). \nonumber \\
\end{eqnarray}
\end{widetext}

Combining the Eqs.~\ref{A:muAmuB}, we obtain the equation for $ \mu_A(\rho,\phi, z) $ 
\begin{widetext}
\begin{eqnarray}\label{A:muAPDE}
-\frac{\mathcal{E}^2-m^2 c^4}{\hbar ^2 c^2}u_A(\rho,\phi, z)
&=&-\left(\frac{\partial ^2}{\partial \rho ^2}+\frac{1}{\rho }\frac{\partial }{\partial \rho }+\frac{1}{\rho ^2}\frac{\partial ^2}{\partial \phi ^2}+\frac{\partial ^2}{\partial z^2}\right)
u_A(\rho,\phi, z), \text{Region I}; \nonumber \\
-\frac{(\mathcal{E}-U)^2-m^2 c^4 }{\hbar ^2 c^2}u_A(\rho,\phi, z)
&=&-\left(\frac{\partial ^2}{\partial \rho ^2}+\frac{1}{\rho }\frac{\partial }{\partial \rho }+\frac{1}{\rho ^2}\frac{\partial ^2}{\partial \phi ^2}+\frac{\partial ^2}{\partial z^2}\right)
u_A(\rho,\phi, z), \text{Region II}. \nonumber \\
\end{eqnarray}
\end{widetext}

We now perform separation of variables for $\mu_A(\rho,\phi, z)$ to let:
\begin{equation}\label{A:muA}
u_A(\rho,\phi,z)=u_A(\rho )e^{i l \phi }\cos(k_m z),
\end{equation}
where the wavefunction $ e^{i l \phi } $ represents a travelling wave along the azimuthal coordinate due to the natural periodic boundary condition. The wavefunction $\cos(k_m z)$ corresponds to a standing wave along the $z$-coordinate due to the strong confinement. Here, $l=0,1,2...$ represents the quantum number of the azimuthal angle and $m=1,3,5...$ corresponds to the quantum number along the $z$-axis for the wave vector:
\begin{equation}
k_m=\frac{m \pi}{2d}.
\end{equation}

The radial wavefunction $u_A(\rho )$ for the spin-up electron can be expressed as: 
\begin{equation}\label{A:muA2}
u_A(\rho )=u(\rho )
\left(
\begin{array}{c}
 1 \\
 0 \\
\end{array}
\right),
\end{equation}

We now derive the ordinary differential equations for $ u(\rho ) $ by substituting Eqs.~\ref{A:muA} and~\ref{A:muA2} into Eq.~\ref{A:muAPDE}:
\begin{widetext}
\begin{eqnarray}\label{A:muAPDF2}
-\frac{\mathcal{E}^2-m^2 c^4}{\hbar ^2 c^2}u(\rho )
&=&-\left( k_m^2+\frac{\partial ^2}{\partial \rho ^2}+\frac{1}{\rho }\frac{\partial }{\partial \rho }-\frac{l^2}{\rho ^2}\right)
u(\rho ), \text{Region I}; \nonumber \\
-\frac{(\mathcal{E}-U)^2-m^2 c^4 }{\hbar ^2 c^2}u(\rho )
&=&-\left( k_m^2+\frac{\partial ^2}{\partial \rho ^2}+\frac{1}{\rho }\frac{\partial }{\partial \rho }-\frac{l^2}{\rho ^2}\right)
u(\rho ), \text{Region II}. \nonumber \\
\end{eqnarray}
\end{widetext}

The analytical solution for $u(\rho )$ is obtained from the above equations:
\begin{equation}\label{A:mu}
u(\rho )=\Biggl\{
\begin{array}{cc}
J_l(\zeta_{nlm}  \rho ) , \text{Region I}\\\\
K_l(\xi_{nlm} \rho), \text{Region II}.\\
\end{array}
\end{equation}
where the functions $ J_l(\zeta_{nlm} \rho) $ and $ K_l(\xi_{nlm} \rho) $ correspond to the Bessel function of the first kind and the modified Bessel function of the second kind, respectively.
The corresponding wave vectors $\zeta_{nlm}$ and $\xi_{nlm}$ are explicitly dependent on the azimuthal and $z$-axis quantum numbers $l$ and $m$ due to the expressions in Eq.~\ref{A:mu}, but implicitly dependent on the radial quantum number $n$ to be determined by the boundary condition. $\zeta_{nlm}$ and $\xi_{nlm}$ are also related to the eigenenergy $\mathcal{E}_{nlm}$:
\begin{eqnarray}\label{A:zetaxi}
\zeta_{nlm}^2&=&\frac{\mathcal{E}_{nlm}^2-m^2 c^4-\hbar ^2 c^2 k_m^2}{\hbar ^2 c^2}; \nonumber \\
\xi_{nlm}^2&=&\frac{(\mathcal{E}_{nlm}-U)^2-m^2 c^4-\hbar ^2 c^2 k_m^2}{\hbar ^2 c^2 }. \nonumber \\
\end{eqnarray}

The large component wave function $u_A(\rho,\phi,z)$ is obtained by combining the solutions: 
\begin{equation}\label{A:muA}
u_A(\rho,\phi,z)=\Biggl\{
\begin{array}{cc}
e^{i l \phi }
\left(
\begin{array}{c}
 J_l(\zeta_{nlm} \rho )\cos(k_m z) \\
 0 \\
\end{array}
\right), \text{Region I} \\\\
\kappa e^{i l \phi }
\left(
\begin{array}{c}
 K_l(\xi_{nlm} \rho)\cos(k_m z) \\
 0 \\
\end{array}
\right), \text{Region II}\\
\end{array}, \\
\end{equation}
where $ \kappa$ is the coupling factor for the wavefunction in Region II. 

The small component wavefunction $ u_B(\rho,\phi,z) $ is subsequently obtained by applying the equations from Eqs.~\ref{A:muAmuB}:
\begin{widetext}
\begin{equation}\label{A:muB}
u_B(\rho,\phi,z)=\Biggl\{
\begin{array}{cc}
e^{i l \phi }
\left(
\begin{array}{c}
-i \eta_{I} k_m J_l(\zeta_{nlm}  \rho )\sin(k_m z) \\
i e^{i \phi}\eta_{I} \{\frac{\zeta_{nlm}}{2}\left[ J_{l-1}(\zeta_{nlm}  \rho )-J_{l+1}(\zeta_{nlm}  \rho )\right]-\frac{l}{\rho } J_l(\zeta_{nlm}  \rho )\}\cos(k_m z) \\
\end{array}
\right), \text{Region I}\\\\
\kappa e^{i l \phi }
\left(
\begin{array}{c}
 -i \eta_{II} k_m K_l(\xi_{nlm}  \rho )\sin(k_m z) \\
i e^{i \phi}\eta_{II} \{\frac{\xi_{nlm}}{2}\left[-K_{l-1}(\xi_{nlm} \rho)-K_{l+1}(\xi_{nlm} \rho)\right]-\frac{l}{\rho } K_l(\xi_{nlm} \rho)\}\cos(k_m z)  \\
\end{array}
\right), \text{Region II}\\
\end{array}
\end{equation}
\end{widetext}
where $ \eta _I $ and $ \eta _II $ represent the geometric factors: 
\begin{eqnarray}
\eta _I&=&\frac{\hbar c}{\mathcal{E}+m c^2};\nonumber\\
\eta _{II}&=&\frac{\hbar c}{\mathcal{E}-U+m c^2},
\end{eqnarray}
which approximately become:
\begin{equation}\label{A:etaapprox}
\eta _{I} \approx \eta _{II} \approx \eta=\frac{\hbar c}{2m c^2}=\frac{\hbar}{2mc},
\end{equation}
due to the eigenenergy $ \mathcal{E}\approx m c^2$ for the confined electron within the quantum well of our concern. 

We have now obtained the eigenwavefunction solution for the spin-up electron within the cylindrical cavity:
\begin{widetext}
\begin{equation}\label{A:psinlm}
\psi_{nlm\uparrow}(\rho,\phi,z)=\Biggg\{
\begin{array}{cc}e^{il\phi} N \left(
\begin{array}{c}
J_l(\zeta_{nlm}  \rho )\cos(k_m z) \\
 0 \\
-i \eta k_m J_l(\zeta_{nlm}  \rho )\sin(k_m z) \\
i e^{i \phi}\eta \{\frac{\zeta_{nlm}}{2}\left[ J_{l-1}(\zeta_{nlm}  \rho )-J_{l+1}(\zeta_{nlm}  \rho )\right]-\frac{l}{\rho } J_l(\zeta_{nlm}  \rho )\}\cos(k_m z) \\
\end{array}
\right), & \text{Region I}\\\\
\kappa e^{il\phi} N \left(
\begin{array}{c}
K_l(\xi_{nlm} \rho)\cos(k_m z) \\
 0 \\
-i \eta k_m K_l(\xi_{nlm}  \rho )\sin(k_m z) \\
i e^{i \phi}\eta \{\frac{\xi_{nlm}}{2}\left[-K_{l-1}(\xi_{nlm} \rho)-K_{l+1}(\xi_{nlm} \rho)\right]-\frac{l}{\rho } K_l(\xi_{nlm} \rho)\}\cos(k_m z)  \\
\end{array}
\right),  & \text{Region 1I}
\end{array}
\end{equation}
\end{widetext}

As an example, we focus our study on the lowest quantum state with $l=0$ and $m=1$. The eigenwavefunction for the spin-up electron is now expressed as:
\begin{widetext}
\begin{equation}\label{A:psin01}
\psi_{n01\uparrow}(\rho,\phi,z)=\Biggg\{
\begin{array}{cc}N\left(
\begin{array}{c}
J_0(\zeta_{n01}  \rho )\cos(k z) \\
 0 \\
-i \eta k J_0(\zeta_{n01}  \rho )\sin(k z) \\
-i e^{i \phi}\eta \zeta_{n01} J_1(\zeta_{n01}  \rho )J_1(\zeta_{n01}  \rho )\cos(k z) \\
\end{array}
\right)  & \text{Region I}\\\\
\kappa N\left(
\begin{array}{c}
K_0(\xi_{n01} \rho)\cos(k z) \\
 0 \\
-i \eta k K_0(\xi_{n01}  \rho )\sin(k z) \\
-i e^{i \phi}\eta \xi_{n01} K_1(\xi_{n01} \rho)K_l(\xi_{n01} \rho)\cos(kz)  \\
\end{array}
\right)  & \text{Region 1I}
\end{array}
\end{equation}
\end{widetext}
where we set $ k_1=k=\frac{\pi}{2d} $. 

We now apply boundary conditions at $ \rho=R $:
\begin{eqnarray}\label{A:boundary}
J_0(\zeta_{n01} R)&=&\kappa K_0(\xi_{n01}  R);\nonumber \\
\zeta_{n01} J_1(\zeta_{n01} R)&=& \kappa \xi_{n01} K_1(\xi_{n01} R).
\end{eqnarray}

By combining the equations from Eqs.~\ref{boundary}, we obtain the following equations for the eigenenergy and the corresponding coupling constant:
\begin{eqnarray}\label{A:boundary2}
\zeta_{n01} K_0(\xi_{n01}  R) J_1(\zeta_{n01} R)&=&\xi_{n01} J_0(\zeta_{n01} R) K_1(\xi_{n01} R);\nonumber \\
\kappa&=&K_0(\xi_{n01}  R)/J_0(\zeta_{n01} R), 
\end{eqnarray}
which gives the eigenenergy $\mathcal{E}_{n01}$ by using the expressions from Eqs.~\ref{A:zetaxi}.


\bibliography{Spin}

\end{document}